\documentclass[superscriptaddress,aps,prx,reprint,nofootinbib,notitlepage,10pt,longbibliography]{revtex4-2}
\usepackage{amsmath}

\usepackage[T1]{fontenc}
\usepackage{amssymb}
\usepackage{amsthm}
\usepackage{mathtools}
\usepackage{braket}

\usepackage{graphicx}
\usepackage{enumitem}
\usepackage{setspace}
\usepackage[autostyle, english = american]{csquotes}
\MakeOuterQuote{"}
\usepackage{hyperref}

\DeclareMathOperator{\Tr}{Tr}

\hypersetup{
    colorlinks=true,
    linkcolor=blue,
    citecolor=blue,
    filecolor=magenta,
    urlcolor=cyan,
}

\begin{document}

% Title information
\title{Energy-Scaled Zero-Noise Extrapolation for Gottesman-Kitaev-Preskill Code}
\author{Gui-Zhong Luo}
 \email{gluo25@wisc.edu}
 \affiliation{Department of Physics, University of Wisconsin -- Madison, Madison, WI 53706, USA}
\author{Matthew Otten}
 \email{mjotten@wisc.edu}
  \affiliation{Department of Physics, University of Wisconsin -- Madison, Madison, WI 53706, USA}
 \affiliation{Department of Chemistry, University of Wisconsin -- Madison, Madison, WI 53706, USA}

\date{\today}

\begin{abstract}
    The performance of Gottesman-Kitaev-Preskill (GKP) codes, an approach to hardware-efficient quantum error correction, is limited by the finite squeezing capabilities of current experimental platforms. To circumvent this hardware demand, we introduce Energy-Scaled Zero-Noise Extrapolation (ES-ZNE), a quantum error mitigation protocol that uses the mean photon number of the GKP code as a tunable effective noise parameter. The protocol measures logical observables at a series of accessible finite energies and extrapolates the results to the ideal, infinite-energy limit using an ansatz based on the code's asymptotic error scaling. Through simulating a GKP qubit under a pure-loss channel, we demonstrate that ES-ZNE successfully mitigates finite-energy errors, recovering the ideal expectation values (within numerical uncertainty) in the shallow-noise regime. Furthermore, by computationally removing artifacts arising from the finite-energy encoding, our method characterizes the intrinsic performance of the ideal GKP code, revealing a sharp error threshold beyond which the code's corrective power diminishes. These results establish ES-ZNE as a practical, software-based strategy for enhancing the performance of near-term bosonic quantum processors, trading sampling overhead for demanding physical resources like high squeezing.
\end{abstract}
\maketitle

\section{Introduction}
\begin{figure*}[!tb] 
    \centering
    \includegraphics[width=\textwidth]{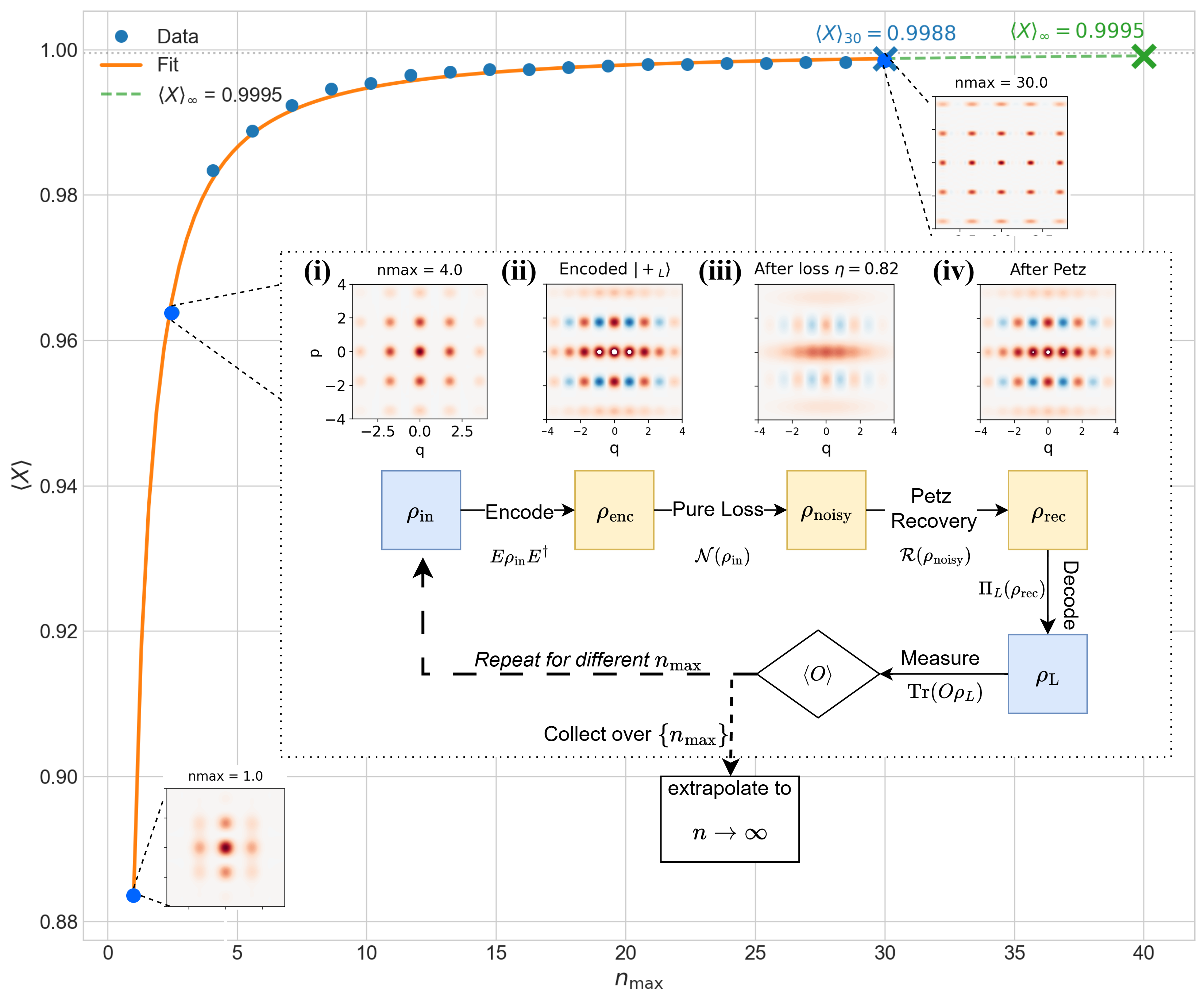}
   \caption{\textbf{Visualization of the extrapolated results and pipeline.} 
\textbf{Extrapolation}: Power-law extrapolation of expectation value 
$\langle X \rangle (n_{\rm max})$, with zoomed-in Wigner functions of the GKP lattice at 
$n_{\rm max} = 1, 4, 30$. 
\textbf{State Evolution ($n_{\rm max}=4$)}: illustration of the GKP code state evolution. 
From left to right: (i) The maximally mixed state of the codespace, $\tfrac{1}{2}P_L$, 
showing the characteristic grid structure. (ii) The encoded logical $|+\rangle_L$ state, 
exhibiting interference fringes indicative of quantum superposition. 
(iii) The state after a pure-loss channel with transmissivity $\eta=0.82$, showing contraction 
towards the origin and decoherence. (iv) The state after the application of the Petz recovery map, 
where the grid structure and interference fringes are partially restored. 
\textbf{Simulation Pipeline:} A schematic of the operational workflow, detailing the transitions 
from a logical input state ($\rho_{\rm in}$) to a physical encoded state ($\rho_{\rm enc}$), 
through the noisy channel ($\rho_{\rm noisy}$) and recovery ($\rho_{\rm rec}$), to the final decoded 
logical state ($\rho_L$) upon which observables are measured. Repeat the pipeline by tuning 
$n_{\rm max}$, and extrapolate to $n_{\rm max} \rightarrow \infty$.}

    \label{fig:main}
\end{figure*}
Quantum computers promise to revolutionize many fields, as demonstrated by recent NISQ-scale experiments in quantum dynamics, chemistry, and materials \cite{GoogleQuantum2025,Alam2025,gratsea2025achieving,nguyen2024quantum}, to name a few. However, current devices remain highly susceptible to noise and errors.
The realization of a large-scale, fault-tolerant quantum computer hinges on our ability to combat noise in quantum systems, a task addressed by the field of quantum error correction (QEC) \cite{Nielsen2000,Cai2023}. While many leading QEC strategies, such as the surface code, offer a path to fault tolerance, they often demand a significant overhead in physical qubits, typically requiring hundreds to thousands of physical qubits per logical qubit to reach algorithmically relevant error rates \cite{Fowler2012,Campbell2017,Babbush2021}.
 Bosonic codes present a hardware-efficient alternative by encoding logical information within the vast Hilbert space of a single harmonic oscillator \cite{Grimsmo2021}. Among these, the Gottesman-Kitaev-Preskill (GKP) code is a particularly promising candidate, offering inherent resilience against small phase-space displacement errors, the bosonic equivalent of bit- and phase-flips \cite{Gottesman2001,CampagneIbarcq2020,Glancy2006}.

%%%%%%edits validating loss rate regime

Bosonic QEC experiments to date primarily benchmark their performance as \emph{quantum memories}: the goal is to extend the lifetime of an encoded
qubit beyond that of the best physical degree of freedom and to approximate the logical identity channel for as long as possible.
Early cat-code demonstrations \cite{Ofek2016,Lescanne2020,Gertler2021} and binomial-code experiments \cite{Hu2019} focused on suppressing amplitude damping of a cavity mode and reported enhanced logical lifetimes.
Subsequent grid-code experiments in superconducting circuits and trapped
ions \cite{CampagneIbarcq2020,deNeeve2022,Sivak2023} adopt the same
perspective: they implement repeated rounds of error correction or
stabilization and quantify performance by the decay of logical Pauli expectation values or average channel fidelity as a function of storage time. In these platforms, the dominant decoherence mechanism of the
bosonic mode is photon loss, with energy-relaxation times
$T_1$ in the $0.1$--$2~\mathrm{ms}$ range and QEC or feedback cycles
in the $1$--$10~\mu\mathrm{s}$ range, so that the per-cycle loss depth
$x_{\text{cycle}} \equiv t_{\text{cycle}}/T_1$ is typically of order $10^{-2}$ \cite{Ofek2016,Hu2019,Lescanne2020,Gertler2021,CampagneIbarcq2020,deNeeve2022,Sivak2023,Terhal2020}.
In particular, Sivak \emph{et al.}\ report $T_1^c \approx 610~\mu\mathrm{s}$ and a grid-code QEC cycle duration $t_c\approx 9.8~\mu\mathrm{s}$, corresponding to $x_{\text{cycle}}\approx 0.016$ per round \cite{Sivak2023}.  Our work is framed in this same memory-oriented setting: we study how well a finite-energy GKP encoding preserves the logical
identity channel under accumulated pure loss, and how this performance can be improved.

%%%%
For the GKP code, error correction capability is linked to the mean photon number (energy) of the encoded state; higher energy provides greater resilience to noise \cite{Albert2018}. However, preparing stable high-energy GKP states is experimentally challenging, requiring high squeezing and complex real-time feedback protocols that push the limits of current hardware \cite{CampagneIbarcq2020}. This difficulty is a primary bottleneck, as fault-tolerant architectures based on GKP codes are predicted to require squeezing thresholds that are at the edge of current experimental capabilities \cite{NohChamberlandBrandao2022,Fukui2018,Menicucci2014}. This trade-off motivates software-based methods to achieve the benefits of high-energy states without the physical cost. Quantum error mitigation (QEM) offers a suite of techniques designed to improve the accuracy of noisy, near-term quantum processors \cite{Cai2023}. In the foreseeable early-fault-tolerant regime, it is anticipated that practical architectures will combine modest-distance QEC with quantum error mitigation techniques such as zero-noise extrapolation, rather than relying on QEC alone \cite{Aharonov2025,Zhong2025,Suzuki2022}.

A prominent QEM protocol is Zero-Noise Extrapolation (ZNE), which estimates the ideal value of an observable by measuring it at several controllably amplified noise levels and extrapolating the results to the zero-noise limit \cite{Li2017, Temme2017,Otten2019Recovering,Otten2019Accounting}. This procedure requires a tunable "noise knob" to scale the error rate, a task often accomplished in qubit systems by inserting additional gates or stretching the duration of control pulses \cite{GiurgicaTiron2020,He2020, Endo2021, Cai2023,LaRose2022}.

In this work, we introduce and numerically validate Energy-Scaled Zero-Noise Extrapolation (ES-ZNE), an instantiation of the ZNE protocol tailored for the GKP code. We use an intrinsic physical parameter of the code—its mean photon number $\bar{n}$—as a natural and calibrated noise knob, where decreasing the energy controllably increases the logical error rate. This approach is conceptually analogous to recent proposals for applying ZNE to logical qubits in surface codes, where the code distance is used as a tunable parameter to scale the logical error rate \cite{wahl2023}. Our protocol involves taking measurements of observables at a sequence of decreasing energies and fitting the resulting expectation values to a numerically motivated power-law model to extrapolate to the infinite-energy ($\bar{n} \to \infty$) limit. We demonstrate that ES-ZNE successfully mitigates artifacts arising from finite-energy encoding, allowing us to computationally remove errors in measurement and characterize the code's fundamental performance. 
In all simulations we take pure loss as the dominant noise channel on the
bosonic mode and choose loss depths $x\in\{0.2,0.4\}$ and mean photon
numbers $\bar n\in[1,30]$.
Anchoring to recent grid-code and cat-/binomial-code experiments,
these values correspond to moderate storage times---a fraction
$\mathcal{O}(0.1$--$0.4)$ of the oscillator lifetime, i.e.\ tens of QEC or
feedback cycles---and an energy range that spans the current few-photon
regime and extends into a higher-energy, more idealized GKP limit
\cite{Ofek2016,Hu2019,Lescanne2020,Gertler2021,CampagneIbarcq2020,deNeeve2022,Sivak2023,Terhal2020,Royer2020}.
Note that throughout this work, the “zero-noise” limit refers to the
ideal logical channel obtained in the $\bar n \to \infty$ limit at fixed physical loss depth $x = -\ln\eta$, rather than to vanishing physical noise in the bosonic mode.

This paper is structured as follows: Section~\ref{sec:background} establishes the theoretical background for the GKP code and the noise and recovery channel. We further propose the ES-ZNE protocol with simulation pipeline in Section~\ref{sec:methodology}. Section~\ref{sec:single_qubit} and \ref{sec:two_qubit} presents our primary numerical results for single- and two-qubit systems. Finally, in Section~\ref{sec:conclusion}, we discuss the implications of our findings and conclude.
%%%%%%%%%%%%%%%%%%%
%%%%%%%%%%%%%%%%%%%

\section{Theoretical Background}
\label{sec:background}
This section establishes the theoretical framework and methodological choices for our numerical study. We define the GKP code within a bosonic mode, outline the physical process of noise and recovery, justify our choice of a near-optimal decoder, motivate the energy parameterization and performance metrics, and define the ES-ZNE protocol used throughout this work.

Fig.~\ref{fig:main} provides a visual summary of the simulation and extrapolation pipeline for a single bosonic mode. The overarching curve shows the extrapolation result based on an evenly-spaced mean photon number. We present a Wigner function plot for mean photon number ($n_{\rm max},$ as defined in Sec~\ref{sec:energy-param}) of $1,4,30$, and, in particular, use the $n_{\rm max}=4$ case to illustrate the pipeline. The top row of the inset illustrates the state's evolution in phase space via Wigner functions, from the initial codespace to the final recovered state. The bottom row presents a schematic of the corresponding operational workflow, from logical input to the measurement of logical observables.

%%%%%%%%%%%%%%%%%%%
\subsection{The Gottesman-Kitaev-Preskill Code in a Bosonic Mode}
\label{sec:gkp code}
We work with a single bosonic mode, setting $\hbar=1$ and
$\hat a=(\hat q+i\hat p)/\sqrt{2}$, $[\hat q,\hat p]=i$.
The ideal (infinite-energy) square-lattice GKP codespace is the common $+1$
eigenspace of the commuting displacement stabilizers
\begin{align}
S_q &= D_{\sqrt{2\pi}} \;=\; e^{-i\,2\sqrt{\pi}\,\hat p},\\
S_p &= D_{i\sqrt{2\pi}} \;=\; e^{+i\,2\sqrt{\pi}\,\hat q},
\end{align}
where $D_\alpha \equiv e^{\alpha \hat a^\dagger-\alpha^{*}\hat a}$ is the phase-space
displacement \cite{Gottesman2001,Albert2018,Weedbrook2012}.
These translate $\hat q$ and $\hat p$ by $2\sqrt{\pi}$, respectively.
While other GKP families (e.g., scaled symplectically self-dual lattices) can be
capacity-achieving under specific conditions \cite{Zheng2025},
all simulations here use the square lattice. We make the following relevant definitions:

\paragraph{Finite-energy  codewords}
Following Albert \emph{et al.} \cite[Eq.~(7.7b)–(7.8)]{Albert2018}, introduce an
envelope parameter $\Delta>0$ and define lattice centers
\begin{equation}
\alpha_{n_1,n_2}^{(\mu)} \;\coloneqq\; \sqrt{\frac{\pi}{2}}\big[(2n_1+\mu)+i\,n_2\big],
\end{equation}
with $(n_1,n_2\in\mathbb Z,\;\mu\in\{0,1\})$, and coherent states $|\alpha\rangle\equiv D_\alpha|0\rangle$.
Up to normalization, the finite-energy (non-orthogonal) logical states are
\begin{align}
\label{eq:gkp_raw_coh}
\big|\tilde{\phi}_{\mu}(\Delta) \big\rangle \propto \nonumber\\
\sum_{n_1,n_2\in\mathbb Z}
&\exp \Big[-\frac{\pi}{2}\Delta^{2}\big((2n_1+\mu)^2+n_2^2\big)\Big] \\
&\exp\Big[-\,i\,\frac{\pi}{4}(2n_1+\mu)n_2\Big]
\big|\alpha_{n_1,n_2}^{(\mu)}\big\rangle . \nonumber
\end{align}
As $\Delta\to0$ these approach the ideal Dirac-comb codewords; for any finite
$\Delta$ the pair $\{\big|\tilde{\phi}_0\big\rangle,\big|\tilde{\phi}_1\big\rangle\}$
is not orthogonal.
This coherent-state construction is equivalent, up to normalization, to
applying a Fock-space envelope operator of the form
$N_\Delta = \exp(-\Delta^2 \hat a^\dagger \hat a)$ to the ideal
infinite-energy grid states, a widely used approach in the finite-energy
GKP literature \cite{Royer2020,Terhal2020,Terhal2016}.
In this representation, the mean photon number scales
approximately as $\bar n \sim 1/(2\Delta^2)$ for small $\Delta$, so that
decreasing $\Delta$ (narrower peaks and wider support in phase space)
corresponds to higher-energy, more ideal codewords
\cite{Royer2020,Terhal2020}.

\paragraph{Codespace projector.}
Let $|\tilde{\phi}_0(\Delta)\rangle,|\tilde{\phi}_1(\Delta)\rangle$ be the finite-energy
coherent-state grid states in Eq.~\eqref{eq:gkp_raw_coh}. Define their $2\times 2$ Gram matrix
$G_{\mu\nu}\coloneqq \langle \tilde{\phi}_\mu|\tilde{\phi}_\nu\rangle$ (positive definite for any finite $\Delta$),
and let $G^{-1/2}$ denote the positive-definite inverse square root
from the spectral decomposition $G=U\Lambda U^\dagger$,
so $G^{-1/2}=U\Lambda^{-1/2}U^\dagger$.
The L\"owdin-orthonormalized logical states \cite{Lowdin1950} are
\begin{equation}
|\phi_\mu\rangle \;\coloneqq\; \sum_{\nu=0}^{1} |\tilde{\phi}_\nu\rangle\,(G^{-1/2})_{\nu\mu},
\qquad \langle \phi_\mu|\phi_\nu\rangle=\delta_{\mu\nu}.
\label{eq.lowdin}
\end{equation}
The encoding isometry is
\begin{equation}
E \;\coloneqq\; \sum_{\mu=0}^{1} |\phi_\mu\rangle\langle \mu|,\qquad E^\dagger E=I_2,
\label{eq:isometry}
\end{equation}
and the codespace projector is
\begin{equation}
\label{eq:codespace_projector}
P_L \;=\; E E^\dagger \;=\; \sum_{\mu=0}^{1} |\phi_\mu\rangle\langle \phi_\mu|.
\end{equation}

%%%%%%%%%%%%%%%%%%%

\subsection{The Pure-Loss Channel and Petz Recovery}
\label{sec:channel_and_recovery}
The primary noise model considered in this work is the pure-loss channel, 
denoted by $\mathcal{N}_\eta$. 
Photon loss is the dominant decoherence mechanism for high-$Q$ (quality factor) bosonic
modes in circuit QED and trapped-ion platforms, and it is the primary
error channel targeted by existing bosonic QEC implementations
\cite{Ofek2016,Hu2019,Lescanne2020,Gertler2021,CampagneIbarcq2020,deNeeve2022,Sivak2023,Terhal2020}.
From the theoretical perspective, the bosonic pure-loss channel is also
the canonical model for analyzing the performance and capacity of GKP and
other continuous-variable codes \cite{Royer2020,Noh2020,Hastrup2023,Shaw2024,Weedbrook2012}.

This channel models the physical process of 
photon loss to an environment and is characterized by a power transmissivity 
$\eta \in [0, 1]$ or equivalently the loss rate $\gamma \equiv 1 - \eta$. 
We find it convenient to parameterize the channel by the loss depth $x = -\ln(\eta)$. 
In the Fock basis, the channel is described by the Kraus operator expansion \cite{Ivan2011, Albert2018}:

\begin{equation}
\mathcal{N}_\eta(\rho) = \sum_{\ell=0}^\infty E_\ell \rho E_\ell^\dagger,
\label{eq:loss_channel}
\end{equation}
where
\begin{equation}
E_\ell = \left( \frac{\gamma}{1-\gamma} \right)^{\ell/2} \frac{\hat{a}^\ell}{\sqrt{\ell!}} (1-\gamma)^{\hat{n}/2}.
\end{equation}

To reverse the effects of this noise, we employ a recovery map $\mathcal{R}$ tailored to the GKP codespace. Our choice is the Petz (transpose) recovery map~\cite{Petz1988,Barnum2002,Ng2010,Gilyen2022,Li2025}, which is known to be near-optimal for approximate quantum error correction and has been successfully applied in recent analytical treatments of GKP codes under pure loss~\cite{Zheng2025} and proposed for experimental implementation~\cite{Png2025}. 

Let $P_L$ denote the projector onto the GKP codespace and define
$
N_L := \mathcal{N}_\eta(P_L).
$
The Petz recovery of $\mathcal{N}_\eta$ with respect to $P_L$ is the CPTP map
\begin{equation}
\label{eq:petz_abstract}
\mathcal{R}(\rho) 
= P_L\, \mathcal{N}_\eta^\dagger\!\big( N_L^{-1/2}\, \rho\, N_L^{-1/2} \big)\, P_L ,
\end{equation}
where $\mathcal{N}_\eta^\dagger$ is the adjoint channel and $N_L^{-1/2}$ denotes the inverse square root of $N_L$ on its support (i.e.\ the positive square root of the Moore--Penrose pseudoinverse of $N_L$).

Writing the loss channel in Kraus form,
\begin{equation}
\mathcal{N}_\eta(\rho) = \sum_{\ell=0}^\infty E_\ell \rho E_\ell^\dagger,
\end{equation}
the Petz map admits the Kraus representation
\begin{equation}
\label{eq:petz_recovery}
R_\ell = P_L E_\ell^\dagger N_L^{-1/2}, \qquad 
\mathcal{R}(\rho) = \sum_\ell R_\ell \rho R_\ell^\dagger .
\end{equation}
For numerical implementation, the potentially ill-conditioned inverse is regularized, e.g.\ by replacing $N_L^{-1/2}$ with $(N_L + \varepsilon I)^{-1/2}$ for a small $\varepsilon > 0$.

%%%%%%

%%%%%%%%%%%%%%%%%%
%%%%%%%%%%%%%%%%%%%
\section{Methodology}
\label{sec:methodology}
%%%%%%%%%%%%%%%%%%%
\subsection{Simulation Pipeline}
\label{sec:pipeline}
Our numerical experiment simulates the complete four-stage quantum error correction process for a single logical qubit encoded in a bosonic mode. This pipeline, visualized in Fig.~\ref{fig:main}, transforms an initial logical state $\rho_{\rm in}$ into a final, decoded logical state $\rho_L$ upon which observables are measured. The process was visualized in Fig~\ref{fig:main}, and consists of the following:

\begin{enumerate}[label=\arabic*.]
    \item \textbf{Encoding:} The process begins with a $2\times2$ logical density matrix $\rho_{\rm in}$ (e.g., representing $\ket{+}_L$). This abstract state is physically realized by embedding it into the infinite-dimensional Hilbert space of the oscillator using the encoding isometry $E$ defined in Sec.~\ref{sec:gkp code}. The resulting physical state is $\rho_{\rm enc} = E\rho_{\rm in} E^\dagger$.

    \item \textbf{Noise Application:} The encoded state evolves under the pure-loss channel $\mathcal{N}_\eta$, as defined in Sec.~\ref{sec:channel_and_recovery}. This step simulates the effect of photon loss, yielding the noisy state $\rho_{\rm noisy} = \mathcal{N}_\eta(\rho_{\rm enc})$.

    \item \textbf{Recovery Operation:} We then apply the near-optimal Petz recovery map $\mathcal{R}$ from Eq.~\eqref{eq:petz_recovery}. This quantum channel is specifically constructed to invert the action of the loss channel on the GKP codespace, producing the recovered state $\rho_{\rm rec} = \mathcal{R}(\rho_{\rm noisy})$.

    \item \textbf{Decoding:} Finally, to evaluate logical performance, the recovered physical state $\rho_{\rm rec}$ is projected onto the codespace and mapped back to a $2\times 2$ logical density matrix via the decoding map
\begin{equation}
  \Pi(\rho_{\rm rec}) = E^\dagger P_L \rho_{\rm rec} P_L E.
\end{equation}
We denote $\rho_L := \Pi(\rho_{\rm rec})$. 

\end{enumerate}

In our simulations the pure-loss channel $\mathcal{N}_\eta$ is applied as
a single block with depth $x = -\ln \eta$ between encoding and recovery.
In hardware, photon loss accumulates continuously during gate operations,
idle periods, and QEC cycles.  For an oscillator with energy-relaxation
time $T_1$, an interval of duration $t$ corresponds to a loss depth
$x = t/T_1$.  Thus, a given $x$ can be interpreted as the total loss
accumulated over a storage time $t = xT_1$ or, equivalently, over
$N = x/x_{\text{cycle}}$ QEC cycles of duration
$t_{\text{cycle}}$, where $x_{\text{cycle}} = t_{\text{cycle}}/T_1$.
Using the parameters of Sivak \emph{et al.}\ for a representative
grid-code experiment \cite{Sivak2023} ($T_1^c \approx 610~\mu\mathrm{s}$,
$t_c \approx 9.8~\mu\mathrm{s}$), our simulated depths
$x\in\{0.2,0.4\}$ correspond to storage times of approximately
$0.2T_1^c$ and $0.4T_1^c$ ($\sim 120$ and $240~\mu\mathrm{s}$) or,
equivalently, to the accumulated loss from about $12$ and $25$ QEC
cycles on the cavity.  We therefore interpret $x$ as a measure of
\emph{accumulated} bosonic loss over intermediate storage durations,
rather than as the noise of a single elementary gate.

%%%%%

%%%%%

%%%%%%%%%%%%%%%%%%%
\subsection{Energy-Scaled ZNE Protocol}
\label{sec:es-zne}
\subsubsection{Rationale and Noise Scaling}
\label{subsec:zne-rationale}
ZNE estimates an observable in the zero-noise limit by measuring it at a sequence of amplified noise levels and extrapolating to the limit of vanishing noise. In our setting, the finite-energy GKP code provides a natural monotone \emph{noise knob}: the target mean photon number $n$ of the codewords. 
As established numerically for the square-lattice GKP code under pure
loss~\cite{Albert2018,Hastrup2023,Shaw2024} and supported analytically
via Petz-based bounds and lattice-geometry arguments
\cite{Royer2020,Terhal2020,Noh2020,Zheng2025}, increasing energy
(decreasing envelope width $\Delta$) monotonically reduces logical error
at fixed loss depth $x = -\ln \eta$ as long as the channel is below its
quantum capacity.

We adopt the near-optimal transpose channel (Petz recovery) framework established in Ref.~\cite{ZhengPRL2024}. This framework relates the code's performance to the properties of the quantum error correction (QEC) matrix. For the specific case of GKP codes under pure loss, Zheng \textit{et al.} recently derived the analytical form of the near-optimal infidelity, $\tilde{\epsilon} :=1 - \tilde{F}^{\text{opt}}$, as a function of the mean photon number $\bar{n}$~\cite{Zheng2025}.

In the asymptotic limit of infinite energy ($\bar n \to \infty$),
the code's near-optimal infidelity $\tilde{\epsilon} := 1 -
\tilde{F}_{\mathrm{opt}}$ is bounded by the geometry of the
symplectic dual lattice $\Lambda^\perp$ \cite{Zheng2025}
\begin{equation}
\lim_{\bar n\to\infty} \tilde{\epsilon}
\;\le\;
\frac{1}{4}
\sum_{\mathbf{x}\in\Lambda^\perp\setminus\{\mathbf{0}\}}
\exp\!\left[
-\pi\,\frac{1-\gamma}{\gamma}\,|\mathbf{x}|^2
\right],
\label{eq:dual-lattice-bound}
\end{equation}
where $\gamma \equiv 1-\eta$ and $|\mathbf{x}|$ is the Euclidean norm.

For a single-mode GKP code based on a square lattice that encodes a
$d_L$-dimensional logical space (so $d_L = 2$ in our simulations),
the shortest nonzero vector of the symplectic dual lattice satisfies
$|\Lambda^\perp|_{\min}^2 = 1/d_L$ \cite{Zheng2025}.  Keeping only the
leading contribution from these shortest vectors, the infinite-energy
infidelity scales as
\begin{equation}
\tilde{\epsilon}_\infty
\;\approx\;
\exp\!\left[
-\frac{\pi}{d_L}\,\frac{1-\gamma}{\gamma}
\right].
\label{eq:square-lattice-scaling}
\end{equation}
This analytic bound provides the theoretical justification for our
extrapolation target:  below the channel capacity threshold, the logical
error vanishes ($\tilde{\epsilon}\to 0$) as the energy diverges.

Crucially, for the finite-energy regime relevant to near-term
experiments ($\bar n \lesssim 30$), the code performance does not
immediately follow this simple exponential scaling.  Instead, the
finite-energy near-optimal infidelity is governed by a more complicated
series of modified Bessel functions.  As derived in
Ref.~\cite{Zheng2025} [Eq.~(G32)], the exact
infidelity of a single-mode GKP code under pure loss can be written
schematically as
\begin{align}
\tilde{\epsilon}(\bar n,\gamma)
=
\frac{1}{d_L} 
& \sum_{\mu,\nu}  \nonumber \\
&\sum_{n_1,n_2,m_1,m_2\in\mathbb{Z}}  
e^{i\phi_{nm\mu\nu}}\,
k^{|L_n|^2 + |L_m|^2}\,
\mathcal{B}_{nm}(t,z),
\label{eq:finite-energy-schematic}
\end{align}
where:
\begin{itemize}
    \item $n_\Delta \equiv 1/(e^{2\Delta^2}-1)$ is the envelope-defined
    energy parameter, and $k = \exp[-\pi(n_\Delta+1/2)]$;
    \item $t = \gamma n_\Delta / (\gamma n_\Delta + 1)$ is a thermal
    factor determined by the loss rate $\gamma$ and $n_\Delta$;
    \item the lattice vectors $L_n$ enumerate a sublattice of the
    symplectic dual lattice $\Lambda^\perp$, as made explicit in
    Eqs.~(21)–(22) of Ref.~\cite{Zheng2025};
    \item the phases $\phi_{nm\mu\nu}$ encode the dual-lattice geometry
    and logical indices (see Eq.~(F29) and surrounding discussion in
    Ref.~\cite{Zheng2025});
    \item $\mathcal{B}_{nm}(t,z)$ is an explicit linear combination of
    modified Bessel functions $I_{\Delta\ell}(z)$ with argument
    $z \propto (n_\Delta+1)\sqrt{t/(1-t)}\,|L_n L_m|$.
\end{itemize}
The full closed form, including all symbol definitions, is provided in
Ref.~\cite{Zheng2025}; we do not reproduce it here in detail as it is not used directly in our numerics.

What matters for our purposes is that $\bar n$ enters
Eq.~\eqref{eq:finite-energy-schematic} only through $n_\Delta$ and the
derived quantities $t$ and $k$, and that for fixed channel parameters
$(\gamma,\eta)$ the resulting $\tilde{\epsilon}(\bar n,\gamma)$
decreases as $\bar n$ increases, in the parameter regime below the
channel capacity \cite{Albert2018,Zheng2025}.  This behaviour, together
with the asymptotic bounds Eqs.~\eqref{eq:dual-lattice-bound}–\eqref{eq:square-lattice-scaling},
confirms that $\bar n$ acts as an effective “noise knob”: increasing the
energy monotonically improves the near-optimal logical performance in
the regimes we consider.
This monotonic relationship in the correctable regime is
consistent with finite-energy GKP analyses in
Refs.~\cite{Royer2020,Terhal2020,Noh2020,Hastrup2023,Vuillot2019} and with the way
experiments parametrize code performance by squeezing or mean photon
number \cite{CampagneIbarcq2020,deNeeve2022,Sivak2023}.

\subsubsection{Extrapolation Models}
\label{subsec:extrap-models}
To estimate the infinite-energy limit of a logical observable from
finite-energy data, we fit its dependence on the mean photon number
$n\equiv\bar n$ using a power-law ansatz:
\begin{equation}
    y(n) = L + c\,n^{-p},
\end{equation}
where $y(n)$ is the measured observable at energy $n$, $L$ is the extrapolated infinite-energy limit, and $p$ is the convergence exponent. 

We note that while the full Bessel-series expression for the finite-energy performance (summarized schematically in Eq.~\eqref{eq:finite-energy-schematic} and given in closed form in Ref.~\cite{Zheng2025}) provides an exact analytical description, utilizing it as a fitting model for ZNE is impractical. The expression involves infinite sums of modified Bessel functions with multiple energy-dependent parameters ($t, k, z$) that are non-trivial to estimate in a black-box experimental setting. Fitting such a complex function to a small set of noisy data points could lead to significant overfitting and numerical instability.

In contrast, we find that the power-law ansatz serves as a robust \textit{effective model} that captures the pre-asymptotic behavior of the Bessel series in Eq.~(\ref{eq:finite-energy-schematic}) within the experimentally relevant regime ($n \le 30$). This choice is a necessary balance between physical motivation and numerical stability for the extrapolation task.

As a diagnostic, we additionally examine residuals on a log--log scale: if $y(n)-L\approx c\,n^{-p}$, then $\log|y(n)-L|$ should be approximately linear in $\log n$ with slope $-p$. This provides a visual check of the model adequacy complementary to goodness-of-fit statistics. Standard Richardson-style polynomial extrapolations in $\lambda\equiv 1/n$ are natural cross-checks in the ZNE literature \cite{wahl2023,Temme2017,Cai2023}. However, such models can be physically unmotivated for large effective noise, where expectation values are expected to decay rather than diverge, potentially leading to instability in the extrapolation \cite{Cai2023}. Our power-law ansatz is chosen to better reflect the expected asymptotic behavior of the system. 
%%%%%%%%%%%%%%%%%%%%%%%%%%%%%%%%

%%%%%%%%%%%%%%%%%%%
\subsection{Numerical Implementation and Metrics}
\label{sec:implementation}

\subsubsection{Energy Parameterization}
\label{sec:energy-param}
The finite-energy GKP codewords are constructed with a Gaussian envelope parameterized by a width $\Delta$. The average photon number of the code, $\bar{n}$, is directly related to this parameter. Following previous GKP analysis~\cite{Zheng2025}, the theoretical energy is given by
\begin{equation}
    \bar{n} = n_\Delta + \delta \approx n_\Delta, \quad \text{where } n_\Delta \equiv \frac{1}{e^{2\Delta^2}-1},
\end{equation}
and $\delta$ represents exponentially small corrections arising from the non-orthogonality of finite-energy states.

In our simulations, we treat the target average photon number,
denoted $\bar n_j$, from the energy schedule
Eq.~\eqref{eq.energy_schedule} as the primary input parameter for each
data point.  To achieve this target, we numerically tune the
corresponding envelope parameter $\Delta_j$.  Specifically, for each
$\bar n_j$ we use a root-finding algorithm to find the value of
$\Delta_j$ that satisfies
\begin{equation}
\frac{1}{d_L}\,\Tr\!\big(\hat{n}\,P_L(\Delta_j)\big)
\;=\;\bar n_j,
\label{eq:delta-calibration}
\end{equation}
with $d_L = 2$ for the single-mode qubit code considered here.  This
ensures that the codespace-averaged mean photon number
$\bar n = d_L^{-1}\Tr(\hat{n} P_L)$ matches the prescribed value.
Throughout, whenever we label a data set by $n_{\rm max}$, this denotes
such a target energy and should be identified with the corresponding
mean photon number, $\bar n_{\rm max}\equiv \bar n_j$.

The resulting energy window $1 \leq \bar n \leq 30$ is chosen to bridge
current experimental capabilities and more idealized GKP encodings.
Existing grid-code and binomial/cat-code experiments in superconducting
circuits and trapped ions typically operate in the few-photon regime,
with $\bar n$ of order unity to $\mathcal{O}(10)$ depending on the
specific finite-energy construction and squeezing level
\cite{Hu2019,Lescanne2020,CampagneIbarcq2020,deNeeve2022,Sivak2023,Terhal2020}.
On the other hand, finite-energy GKP theory and capacity analyses
\cite{Royer2020,Noh2020,Hastrup2023,Shaw2024} commonly consider
higher-energy codewords as a route toward approaching the ideal GKP
limit and the pure-loss channel capacity.  Our simulations therefore
span from near-term experimentally relevant energies to an aspirational
higher-energy regime where the GKP structure is more sharply defined but
still numerically tractable.

\subsubsection{Energy Schedule}
\label{subsec:energy-schedule}
Fix a loss depth $x$ and select a largest feasible energy $n^*$ allowed by the Hilbert-space truncation and computational budget. This establishes the effective noise range. We then construct a decreasing ladder of energies
\begin{equation}
n_0 \;=\; n^* \;>\; n_1 \;>\; \cdots \;>\; n_{K-1},
\end{equation}
and for each $n_j$ we instantiate codewords by tuning the envelope parameter $\Delta$ so that the realized mean photon number satisfies $\bar n \approx n_j$ (see Sec.~\ref{sec:energy-param} for the calibration procedure). 

All simulations in this work use an evenly spaced (arithmetic) energy schedule
\begin{equation}
n_j \;=\; n_0 \;-\; j\,\Delta n, \qquad j=0,1,\dots,K-1,
\label{eq.energy_schedule}
\end{equation}
with a fixed step $\Delta n>0$. For each $n_j$, we run the full pipeline in Sec~\ref{sec:pipeline} and record both leak-aware and conditional logical expectations as shown in the following subsection.

\subsubsection{Reporting Metrics and Uncertainty}
\label{sec:metric}
All simulation outputs are derived from the final logical block $\rho_L$, obtained after the full pipeline described in Sec.~\ref{sec:pipeline}. We report the following metrics, which distinguish between the total expectation value and the value conditioned on remaining within the logical codespace. This conditioning is a form of post-selection, a common strategy in error mitigation protocols \cite{Cai2023}.
\begin{enumerate}[label=\arabic*.]
    \item \textbf{Codespace Survival Probability (Weight):} The trace of the final logical state, which represents the probability of not leaking out of the codespace.
    \begin{equation} 
        w = \Tr (\rho_L) \in [0,1]. 
    \end{equation}
    \item \textbf{Leak-aware and Conditional Expectations:} For any logical Pauli observable $O \in \{X,Y,Z\}$, we report both the leak-aware expectation, $\langle O \rangle_{\text{leak}} = \mathrm{Tr}(O \rho_L)$, and the conditional expectation, which post-selects on survival in the codespace:     
    \begin{equation} 
        \langle O \rangle_{\text{cond}} = \frac{\langle O \rangle_{\text{leak}}}{w} \quad (\text{for } w > 0).
    \end{equation}
\end{enumerate}
We enforce the physical invariants $0 \le w \le 1$ and $|\langle O \rangle_{\text{cond}}| \le 1$ as sanity checks on our numerical implementation.

 Uncertainty intervals for the extrapolated parameters $(L, p)$ from the model in Sec.~\ref{subsec:extrap-models} are estimated using a non-parametric bootstrap procedure. We resample the simulated $(n_j, \langle O \rangle_j)$ data with replacement, repeat the fitting process on each of the bootstrap samples, and report the standard deviation of the resulting parameter distributions as the standard error. This procedure provides error bars for our extrapolated results.

\section{ES-ZNE for single qubit states}
\label{sec:single_qubit}

In this section, we first characterize the performance of the finite-energy GKP code under pure loss to establish a baseline. We then present the primary results of applying our ES-ZNE protocol to mitigate these finite-energy errors. Finally, we will analyze the behavior of the extrapolation across different noise regimes to gain insight into the code's convergence properties. 

\subsection{Baseline Performance of Finite Energy GKP codes}
\begin{figure}[h]
    \centering
    \includegraphics[width=1\linewidth]{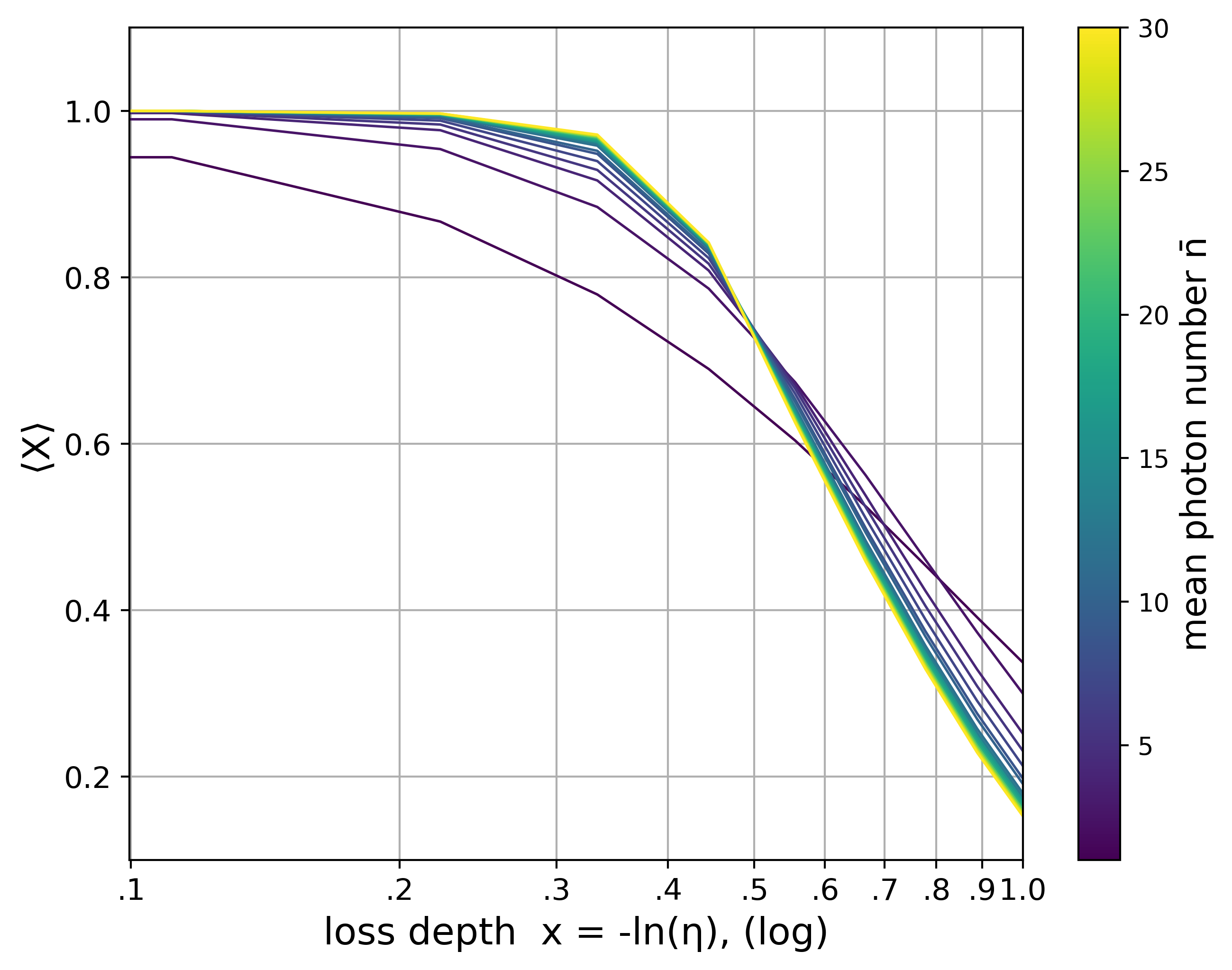}
    \caption{\textbf{Baseline performance of finite-energy GKP codes.} The conditional expectation value $\langle X \rangle$ is plotted against loss depth $x = -\ln(\eta)$ (log scale) for varying mean photon numbers $\bar{n}$ (color bar). The data reveals two distinct regimes separated by a threshold near $x \approx 0.45$. In the low-noise regime, increasing energy (lighter colors) monotonically improves coherence. Conversely, in the high-noise regime, the trend reverses, and higher-energy states degrade more rapidly as the channel capacity is exceeded.}
    \label{fig:coherence}
\end{figure}

To establish and motivate the need for error mitigation, we first analyze the performance of the finite-energy GKP code as a function of both noise strength and available energy. Figure \ref{fig:coherence} shows the conditional expectation value $\langle X \rangle_{\text{cond}}$ for a logical $|+\rangle_L$ state, plotted against the loss depth $x = -\ln(\eta)$ for an even-spaced range of mean photon numbers $\bar{n}$ from 1 to 30.

Three clear trends emerge from the data. First, for any constant mean photon number (i.e. following any single curve), the logical coherence degrades monotonically as the loss depth \( x \) increases. This is the expected behavior, as greater photon loss leads to more significant state contraction and decoherence in phase space, which the Petz recovery map can only partially reverse.

Second, for any fixed loss depth \( x \) (i.e., looking at a vertical slice through the plot), performance systematically improves as the mean photon number \( \bar{n} \) is increased. This trend is consistent with prior analytical and numerical studies of GKP codes under loss \cite{Albert2018, Zheng2025} and confirms the rationale for our energy-scaling approach outlined in Sec~\ref{subsec:zne-rationale},where higher energy corresponds to a wider separation of the GKP lattice peaks, providing greater resilience against the noise. Energy thus acts as a critical resource for improving the code's performance. 

Finally, we notice the appearance of a threshold or crossover region around a loss depth of \( x \approx 0.4\). At this point,  the performance curves for all but the lowest energy codes appear to converge, and the rate of coherence decay steepens significantly. This feature suggests a transition in the code's error-suppression ability. Below this threshold (\( x \lesssim 0.4 \) ), the code and recovery channel are highly effective, and increasing energy provides a clear benefit. Above this threshold, the loss appears to become too severe for the recovery map to effectively correct, leading to a much more rapid decline in logical coherence across all energy levels.
We study the behavior near the theoretical threshold (loss depth $x \approx 0.4$) and in the regime beyond capacity ($x \approx 0.556$) in Fig~\ref{fig:powerlaw_1q}.

Despite the clear improvements of increasing energy below the threshold, a noticeable gap persists between the performance at the highest simulated energy \( \bar{n} = 30 \) and the ideal value of \( \langle X \rangle = 1  \), especially at higher loss depth. This residual error is a direct consequence of the finite energy of the code. The goal of our ES-ZNE protocol is to systematically estimate and remove this finite-energy error by extrapolating the observed trend to the zero noise limit (\( \bar{n}\rightarrow \infty\)  ).

\subsection{Error Mitigation via Energy-scaled Extrapolation}

\begin{figure}[ht]
    \centering
    \includegraphics[width=1\linewidth]{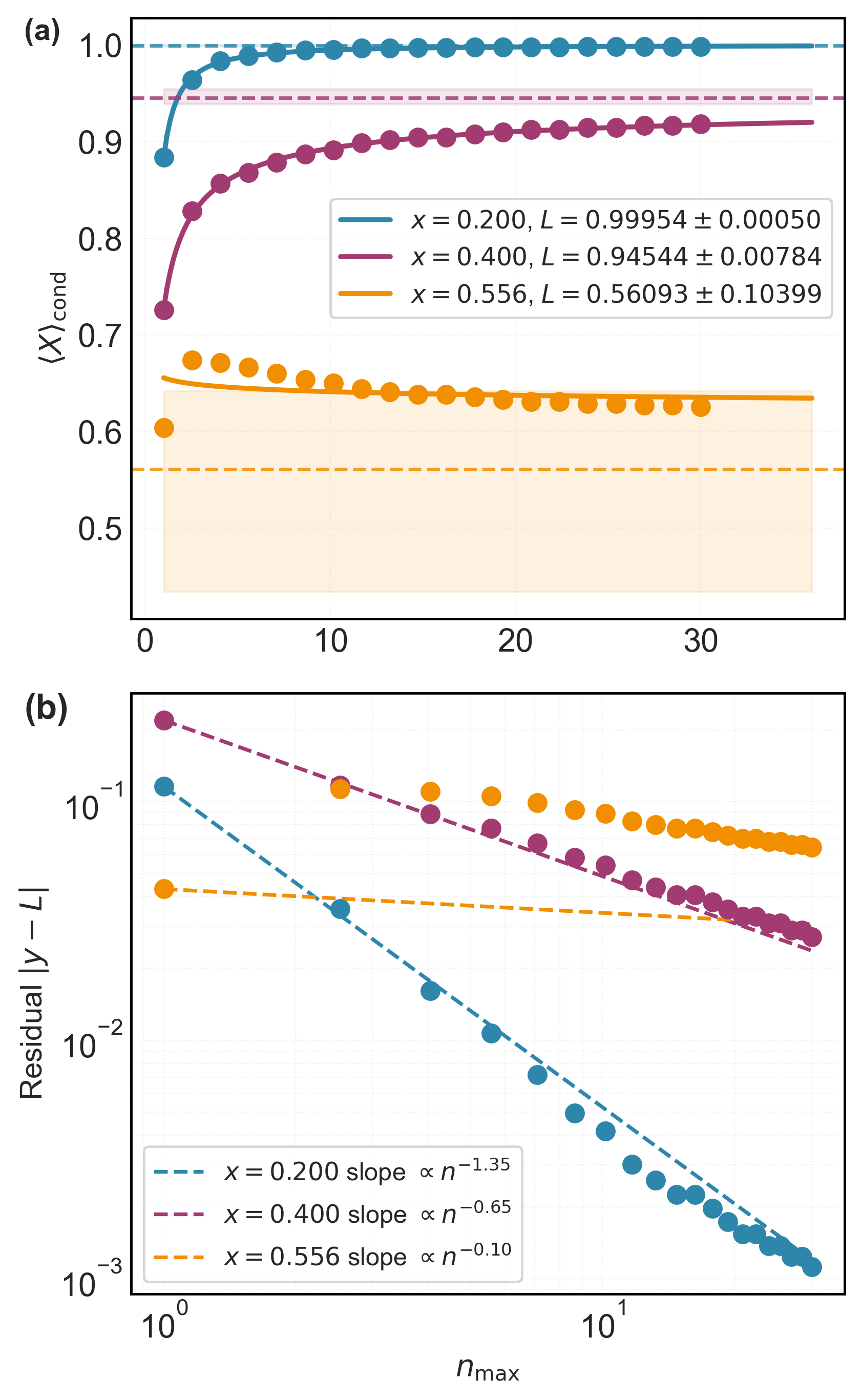}
    \caption{\textbf{Single-Qubit Zero-Noise Extrapolation. }\textbf{(a):} Conditional expectation values $\langle X \rangle_{\text{cond}}$ and power-law extrapolations for three representative loss depths. Shaded regions indicate the uncertainty of the extrapolated limit $L$. For the protected regime ($x=0.2, 0.4$), we observe monotonic improvement with energy. In contrast, the post-threshold regime ($x=0.556$) exhibits performance degradation as energy increases. \textbf{(b):} Residual diagnostic $|y(n) - L|$ on a log-log scale. The datasets for $x=0.2$ and $0.4$ display linear behavior, confirming the validity of the power-law ansatz, whereas the $x=0.556$ data deviates significantly from this scaling.}
    \label{fig:powerlaw_1q}
\end{figure}

We now apply the ES-ZNE protocol described in Sec.~\ref{sec:es-zne} to the data generated by the simulation pipeline. Figure~\ref{fig:powerlaw_1q} illustrates this process and its outcome for three different loss depths. 

We first focus on extrapolation at shallow loss (blue datapoints and curve in Fig.~\ref{fig:powerlaw_1q}).
The power-law model provides an excellent fit to the simulation data, accurately capturing the diminishing returns of increasing the mean photon number. For the low-loss regime, the extrapolation to the infinite-energy limit yields a conditional expectation value of \( \langle X \rangle_\infty = 0.99954 \pm 0.0005. \) This value represents a significant improvement over the best-performing finite-energy datapoint at \( \bar{n} = 30 \), \(\langle X\rangle_{30}=0.9988\). Crucially, the extrapolated result is consistent with the ideal, noiseless value of \( \langle X \rangle = 1 \) within the estimated uncertainty. This provides clear evidence of the successful mitigation of the finite-energy error.

The validity is further supported by the residual analysis, \( | y(n)-L |  \) on a log-log scale. The observed linear trend confirms that the finite energy error decays as a power-law in \( n \), reinforcing our choice of the power-law model and the fidelity of the extrapolated limit.

\begin{figure}[ht]
    \centering
    \includegraphics[width=1\linewidth]{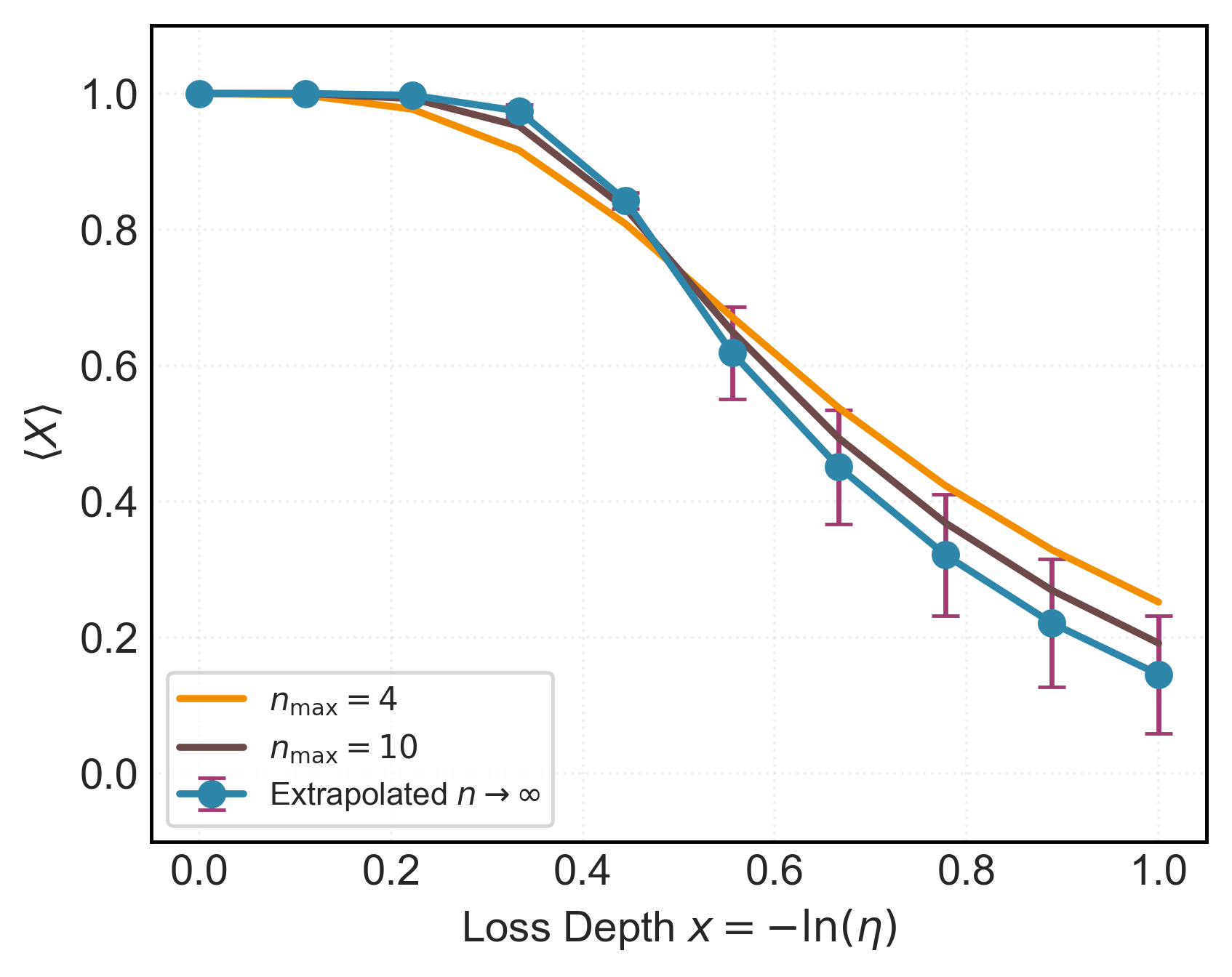}
    \caption{\textbf{Extrapolation across regimes.} Extrapolated infinite-energy limit of \(\langle X \rangle\) against loss depth \(x = -\ln(\eta)\), compared with measurements from finite-energy simulations ($n_{\rm max}=4, 10$). The extrapolated measurement achieves higher expectation with high certainty before the threshold, but performs worse than finite-energy measurements beyond the threshold with higher uncertainty.}
    \label{fig:extrapolation_over_x}
\end{figure}

\subsection{Analysis of Performance Across noise Regimes}
To assess the protocol's robustness and understand how the fundamental performance of the GKP code depends on noise, we now analyze the extrapolated results across the full range of simulated loss depths. Figure \ref{fig:extrapolation_over_x} summarizes our central findings, plotting the infinite-energy limit $L = \langle X \rangle_\infty$ as a function of the loss depth $x$, in comparison with measurements from finite energy simulation. Each data point represents the final output of the ES-ZNE protocol for a given noise level, with the error bars indicating the bootstrap-estimated uncertainty of the extrapolation.
The plot reveals a clear, sigmoidal relationship, characteristic of a threshold behavior in error correction. We can identify three distinct operational regimes:
\paragraph{The Protected Regime ($x \lesssim 0.3$)}
For low loss depths, the extrapolated performance is consistently pinned at $L \approx 1.0$ with negligible uncertainty. This demonstrates that in the ideal infinite-energy limit, the GKP code and Petz recovery can achieve near-perfect correction, completely suppressing the effects of noise in this regime. This is exemplified with the $x=0.2$ data set in Fig.~\ref{fig:powerlaw_1q}.

\paragraph{The Threshold Regime ($0.3 \lesssim x \lesssim 0.5$)}
Around a loss depth of $x \approx 0.4$, the code's performance begins a noticeable, continuous degradation. This transition corresponds to the quantum capacity limit of the pure-loss channel for a rate-1 code. The quantum capacity of the pure-loss bosonic channel with transmissivity $\eta$ is 
$C_Q(\eta) = \max\!\bigl(\log_2 \frac{\eta}{1-\eta}, \bigr)$
~\cite{Holevo2001,Weedbrook2012}. 
In the regime where error-free transmission is possible ($C_Q \ge R$), this capacity limit was recently proven to be achievable by ideal GKP codes (in the infinite-energy and multimode limit)~\cite{Zheng2025}. For any code that encodes one logical qubit per use of the bosonic mode (rate $R=1$ qubit/mode), asymptotically error-free transmission or storage is only possible if $C_Q(\eta)\ge 1$. This condition yields a critical transmissivity $\eta_{\text{crit}} = 2/3$, corresponding to a loss depth
\begin{equation}
x_{\text{crit}} = -\ln(2/3)\approx 0.4055.
\end{equation}
Remarkably, our ES-ZNE protocol identifies the breakdown in correctability around this theoretical limit (Fig.~\ref{fig:extrapolation_over_x}). This finding is consistent with the performance degradation observed in other recent analytical and numerical studies of finite-energy GKP codes under pure loss~\cite{Royer2020,Hastrup2023,Shaw2024}. Below $x_{\text{crit}}$, the pure-loss channel can support at least one reliably transmitted qubit per use, so, consistently, our extrapolated logical expectation values approach the ideal value $L = 1$ in this regime. For $x > x_{\text{crit}}$, the quantum capacity drops below one qubit per use, and even an ideal GKP code of unbounded energy cannot perfectly preserve a single logical qubit.

\paragraph{The Decoherence Regime ($x \gtrsim 0.5$)}
In the high-loss region, the performance curve flattens out at a low value. Crucially, we observe that the measured expectation value \textit{decreases} as we increase the mean photon number (see $x=0.556$ dataset (yellow) in Fig.~\ref{fig:powerlaw_1q}), a reversal of the monotonic improvement observed in the low-loss regime~\cite{Zheng2025}. This behavior arises because we are operating well beyond the capacity threshold ($x > 0.405$). In this regime, the high-energy GKP states possess sharp peaks that are washed out by the noise, causing the Petz recovery map—which attempts to map states back to the logical subspace—to make high-confidence errors that displace the state away from the logical axis. In contrast, lower-energy states are already smeared and effectively thermal, leading to a "confused" but less aggressively erroneous recovery. This signals the fundamental breakdown of the coding scheme and the non-invertibility of the noise channel above the capacity threshold.
Consequently, any QEM protocol would require an increasingly large sampling overhead to be effective, rendering the extrapolation inherently uncertain and unreliable \cite{Cai2023, Takagi2022,Takagi2023}.

In summary, by using the ES-ZNE protocol to systematically remove the implementation-dependent error of finite energy, we have isolated and quantified the fundamental performance limits of the ideal GKP code under pure loss. Figure \ref{fig:extrapolation_over_x} visualizes the phase transition in the code's error-correcting capability, providing a quantitative estimate for its operational threshold and demonstrating that our extrapolation method is robust and numerically self-consistent across all noise regimes.

\section{Expanding to Two Qubits}
\label{sec:two_qubit}
Having established the formalism and effectiveness of the ES-ZNE protocol for a single logical qubit, we now extend our analysis to a two-qubit system. We will first develop the theoretical framework for a two-qubit system under the assumption of independent noise channels, demonstrating how the problem decomposes into single-qubit characterizations. We then define a state-averaged performance test to benchmark the performance of the error correction pipeline in this two-qubit setting.

\subsection{Two-Qubit Framework and Product Channels}

Consider a system of two logical qubits, where each is encoded into a separate, independent bosonic mode, denoted A and B. The two-qubit logical Hilbert space is the direct product of the single-qubit spaces, \( \mathcal{H}_{L} = \mathcal{H}^{(A)}_{L} \otimes \mathcal{H}^{(B)}_{L} \), and the physical space is similarly \( \mathcal{H}_{\text{phys}} = \mathcal{H}_{A} \otimes \mathcal{H}_{B} \).

The encoding of a two-qubit logical state into the physical system is described by the direct product of the single-qubit encoding isometries defined in Eq.~\eqref{eq:isometry}
\begin{align}
    E_{AB} \coloneqq E_{A} \otimes E_{B}.
\end{align}
An arbitrary two-qubit logical input state \( \rho_{\text{in}} \) is thus encoded as \( \rho_{\text{enc}} = E_{AB} \rho_{\text{in}} E^{\dagger}_{AB} \).

We assume the two bosonic modes are subject to independent noise processes. The total noise channel is therefore a product of the single-mode pure-loss channels defined in Eq.~\eqref{eq:loss_channel}
\begin{align}
    \mathcal{N} \coloneqq \mathcal{N}_{\eta_A} \otimes \mathcal{N}_{\eta_B}.
\end{align}
This physical independence motivates the choice of a separable recovery operation, which consists of applying the Petz recovery map, as defined in Eq.~\eqref{eq:petz_recovery}, to each mode individually:
\begin{align}
    \mathcal{R} \coloneqq \mathcal{R}_A \otimes \mathcal{R}_B.
\end{align}

The entire physical process—encoding, noise, and recovery—is described by
an effective map on the two-qubit logical space, $\Lambda_{AB}$. Under our
assumption of separability, this map is a direct product of single-qubit
effective channels, $\Lambda_{AB} = \Lambda_A \otimes \Lambda_B$, where each
$\Lambda_i$ is defined as
\begin{align}
  \Lambda_i(\rho)
  \;=\;
  E_i^\dagger\Big[
    P_L^{(i)}\big(\mathcal{R}_i\circ \mathcal{N}_{\eta_i}\big)
    \big(E_i\,\rho\,E_i^\dagger\big)P_L^{(i)}
  \Big]E_i,
\end{align}
with $E_i$ the encoding isometry and $P_L^{(i)}$ the corresponding
single-mode codespace projector.

The full two-qubit expectation value of an observable
$O_{ab} = \sigma_a \otimes \sigma_b$ for an input state $\rho_{\text{in}}$ is
\begin{align}
    \langle O_{ab} \rangle_{\text{leak}}
    = \Tr_L \!\big[ (\sigma_a \otimes \sigma_b)
       \, (\Lambda_A \otimes \Lambda_B) (\rho_{\text{in}}) \big].
\end{align}
To evaluate this, we expand the input state in the Pauli basis,
\begin{align}
    \rho_{\text{in}}
    = \frac{1}{4} \sum_{\mu, \nu=0}^3
      A_{\mu\nu}\, (\sigma_\mu \otimes \sigma_\nu),
\end{align}
where $A_{\mu\nu} = \Tr[(\sigma_\mu \otimes \sigma_\nu)\rho_{\text{in}}]$ are
the real Pauli coefficients of the input state. Substituting this into the
expectation value and using the properties of the tensor product, we find
\begin{align}
    \langle O_{ab} \rangle_{\text{leak}}
    = \frac{1}{4} \sum_{\mu,\nu=0}^3
      A_{\mu\nu} \,
      \big( \Tr_L[\sigma_a \Lambda_A(\sigma_\mu)] \big)
      \big( \Tr_L[\sigma_b \Lambda_B(\sigma_\nu)] \big).
\end{align}

To express this more compactly, we define the Pauli transfer matrix (PTM) for
a single-qubit channel $\Lambda$. Its real entries are
\begin{align}
    \chi_{ij} \coloneqq \frac{1}{2} \Tr_L[\sigma_i \Lambda(\sigma_j)],
\end{align}
which fully characterize the map $\Lambda$ by describing how it transforms
the Pauli components of an input state~\cite{Gilchrist2005}.
The two-qubit expectation value can then be written as a bilinear contraction
\begin{align}
\langle O_{ab}\rangle_{\text{leak}}
  =  \sum_{\mu,\nu = 0}^{3}
    A_{\mu\nu}\, \chi^{(A)}_{a\mu}\, \chi^{(B)}_{b\nu}.
\end{align}

This decomposition, which reveals that the two-qubit channel is fully characterized by the single-qubit process matrices, is a general feature of separable quantum channels and can be formally described using the PTM formalism or tensor network calculus \cite{Wood2015}. We need only compute four single-qubit maps $\Lambda(\sigma_\mu)$ per mode, reducing computational complexity for simulation purposes. The product structure also implies factorization of survival probability $w_{AB}=w_A w_B$ and conditional expectations for product input states.

Crucially, this framework assumes statistically independent noise and uncorrelated recovery. It excludes correlated loss mechanisms or joint recovery strategies that might exploit them. While these are important future directions, the product-channel assumption provides a physically motivated baseline for spatially separated modes.

We validate this framework in the next section by applying ES-ZNE to a concrete benchmark: the maximally entangled logical $|\Phi^+\rangle_L$ state measured on the $XX$ correlator. This test verifies that our extrapolation methodology extends naturally to entanglement witnesses under the product-channel decomposition.

\subsection{Extrapolation of Two-Qubit Entangled States}

To validate the two-qubit framework and demonstrate the extensibility of our ES-ZNE protocol, we apply it to the maximally entangled logical state $|\Phi^+\rangle_L = \frac{1}{\sqrt{2}} (|0\rangle_L |0\rangle_L + |1\rangle_L |1\rangle_L)$, measured via the two-qubit correlator $\langle XX \rangle$. We assume identical independent pure-loss channels on each mode with loss depths $x \approx 0.2$ (low noise) and $x \approx 0.4$ (moderate noise), mirroring the single-qubit analyses in Fig.~\ref{fig:powerlaw_1q}.

The simulation pipeline follows the product-channel decomposition described above, with encoding, noise, recovery, and decoding applied separately to each mode. We use the same energy schedule (even-spaced ladder from $\bar{n} = 1$ to $\bar{n} = 30$) and power-law extrapolation model as in the single-qubit case (Sec.~\ref{subsec:extrap-models}). This leads to extrapolation behavior that is qualitatively similar to the single-qubit case, with finite-energy errors diminishing as a power-law in $\bar{n}$.

\begin{figure}[h!]
    \centering
    \includegraphics[width=1\linewidth]{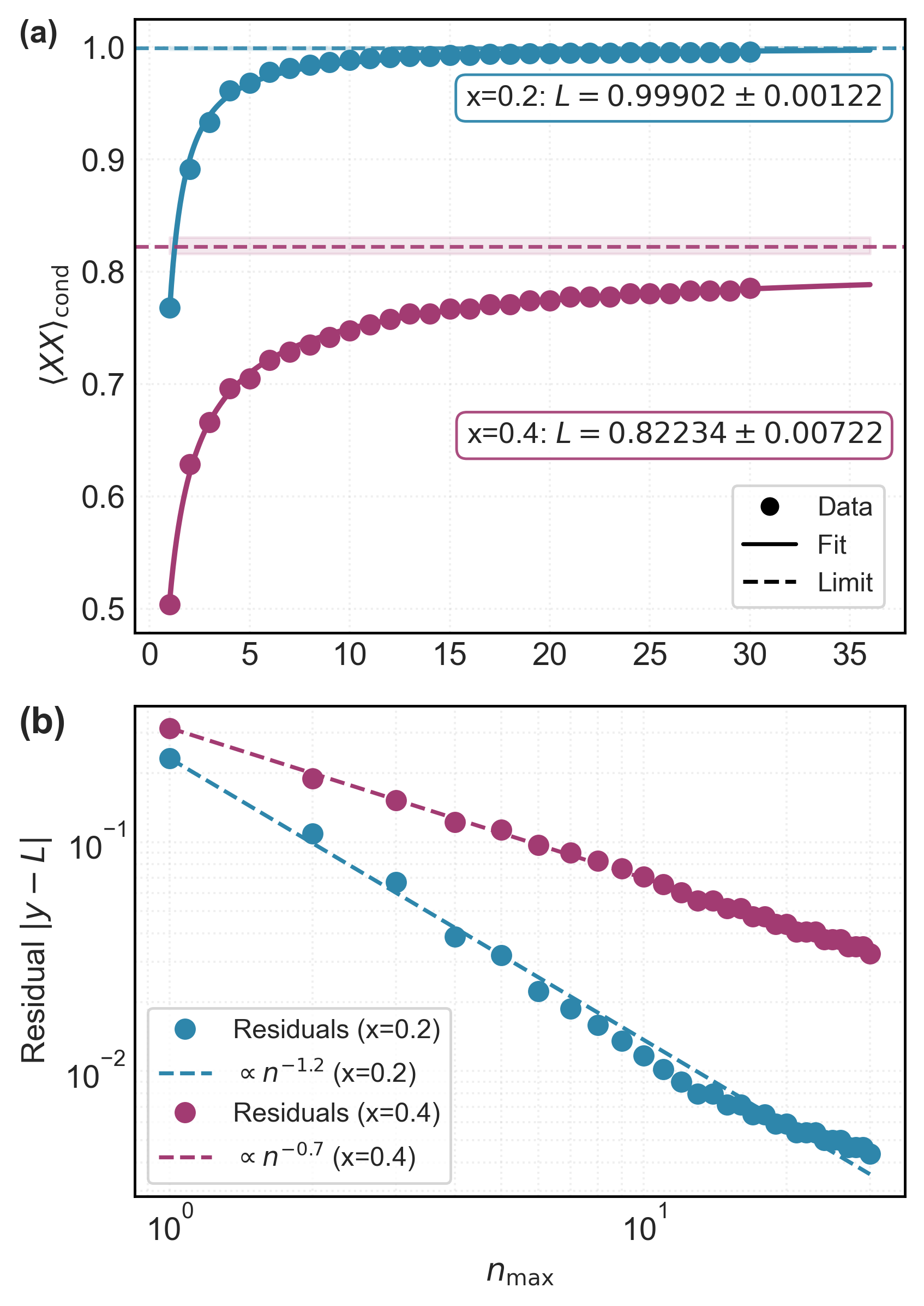} 
    \caption{\textbf{Two-Qubit Zero-Noise Extrapolation}. 
    \textbf{(a):} Power-law extrapolation of the conditional expectation value $\langle XX \rangle_{\text{cond}}$ for the maximally entangled state $|\Phi^+\rangle_L$. Data is shown for independent pure-loss channels at depths $x \approx 0.2$ (blue) and $x \approx 0.4$ (purple). Shaded regions indicate the uncertainty of the extrapolated limit $L$. The low-loss case successfully recovers the ideal correlation ($L \approx 1$), while the higher-loss case reveals residual decoherence ($L \approx 0.82$) consistent with the single-qubit threshold behavior. 
    \textbf{(b):} Residual diagnostic $|y(n) - L|$ on a log-log scale. The linear behavior confirms that the power-law ansatz accurately models finite-energy errors in the two-qubit setting.
    }
    \label{fig:fit_2q}
\end{figure}

Figure~\ref{fig:fit_2q} shows the results for both loss depths. As expected, the power-law fits capture the convergence to the infinite-energy limit, with log-log residual plots confirming the model's adequacy. At low loss ($x \approx 0.2$), the extrapolated value $L = 0.99902 \pm 0.00122$ is consistent with the ideal $\langle XX \rangle = 1$ within uncertainty, demonstrating effective mitigation of finite-energy errors. At moderate loss ($x \approx 0.4$), the extrapolated value $L = 0.82234 \pm 0.00722$ reflects partial decoherence, similar to the single-qubit threshold onset, but with slightly larger uncertainty due to the compounded effects across two modes.

These results confirm that ES-ZNE extends naturally to multi-qubit settings under independent noise, yielding reliable infinite-energy estimates. However, we note that the extrapolated performance is not guaranteed to reach unity beyond the single-qubit threshold (as seen in Fig.~\ref{fig:extrapolation_over_x}), and correlated noise could alter this behavior—future work should investigate such extensions.

\subsection{Coherence with Randomly-Generated Two-Qubit States}

To further assess the robustness of the GKP code and ES-ZNE protocol in preserving two-qubit correlations under pure loss, we introduce a protocol that averages the logical coherence error over an ensemble of randomly generated input states. This state-averaged approach provides a more comprehensive measure of performance than tests on a few fixed states (such as $|\Phi^+\rangle_L$), as it assesses the pipeline's robustness across the entire logical Hilbert space.

For each of 50 independent trials, we sample a two-qubit pure state $|\psi\rangle$ from the Haar measure on $\mathbb{C}^4$ by normalizing a vector of complex random numbers drawn from a standard normal distribution \cite{Mezzadri2007}, and construct the logical density matrix $\rho_{AB} = |\psi\rangle\langle\psi|$. We then compute its Pauli decomposition coefficients $A_{\mu\nu} = \Tr[(\sigma_\mu \otimes \sigma_\nu) \rho_{AB}]$. For each energy level $n_{\max}$ in the schedule (Eq.~\eqref{eq.energy_schedule}) and each loss depth $x$, we evaluate the conditional expectations $\langle O \rangle_{\text{cond}}(n_{\max}, x)$ for the observable set $S = \{XX, YY, ZZ\}$ using the product-channel pipeline and cached single-qubit responses (as per the decomposition in the previous subsection).

The ideal (noise-free) expectations $\langle O \rangle_{\text{ideal}}$ are approximated by the conditional expectations at a reference depth of $x=0$, where no loss occurs. This choice is justified because the same random logical state is used for both the noisy and reference cases, isolating deviations due solely to the encode–loss–Petz–decode process.

For each trial, we compute a per-trial error metric
\begin{equation}
\Delta_E^{\text{trial}}(n_{\max}, x) = \frac{1}{|S|} \sum_{O \in S } \big| \langle O \rangle_{\text{cond}}(n_{\max}, x) - \langle O \rangle_{\text{ideal}} \big|.
\end{equation}
Averaging over all trials yields the final metric $\Delta_E(n_{\max}, x)$, which quantifies the state-averaged (via Haar sampling) and observable-averaged absolute error in recovered logical correlators relative to the ideal. Physically, $\Delta_E$ measures the fidelity of two-qubit Pauli correlation preservation after pure-loss noise of strength $x$, highlighting how increasing $n_{\max}$ (i.e., higher-energy codewords) enhances coherence retention under the GKP code and Petz recovery.

\begin{figure}[h!]
    \centering
\includegraphics[width=1\linewidth]{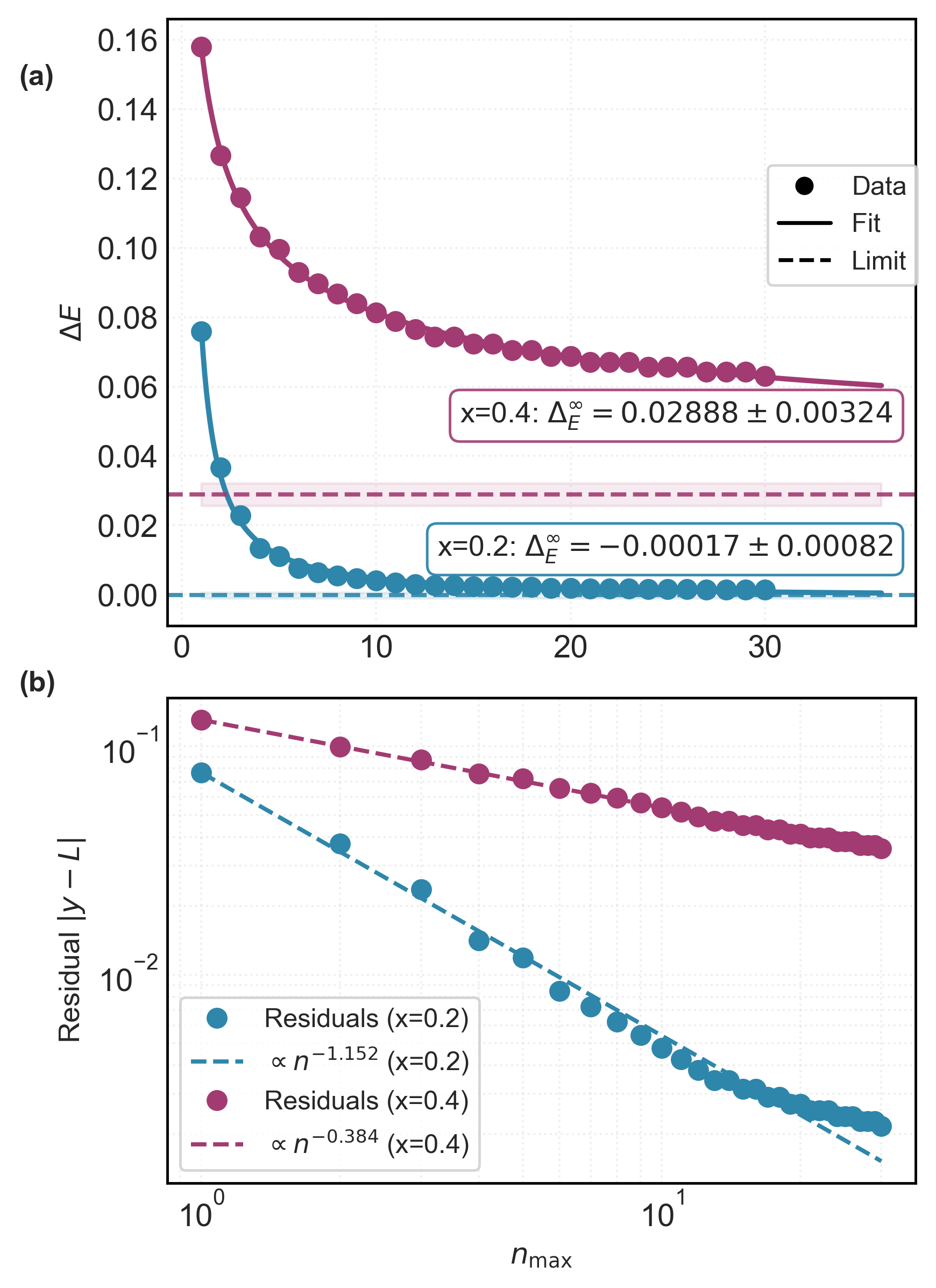}  
    \caption{\textbf{State-averaged Two-Qubit Extrapolation.}
    \textbf{(a):} Power-law extrapolation of the mean error metric $\Delta_E$, averaged over an ensemble of Haar-random two-qubit input states. Data is shown for independent pure-loss channels at depths $x=0.2$ (blue) and $x=0.4$ (purple). Shaded regions indicate the uncertainty of the extrapolated limit $\Delta_E^\infty$. At low loss, the extrapolated error vanishes ($\Delta_E^\infty \approx 0$), indicating that the protocol preserves coherence across the entire logical Hilbert space. At higher loss, a finite residual error persists, mirroring the threshold behavior observed in specific eigenstates. 
    \textbf{(b):} Residual diagnostic $|y(n) - L|$ on a log-log scale. The linearity confirms that the finite-energy error decays as a power law even for this state-averaged metric.
    }
    \label{fig:fit_2q_rand}
\end{figure}

We apply the same power-law extrapolation (Sec.~\ref{subsec:extrap-models}) to $\Delta_E$ as a function of $n_{\max}$ for fixed loss depths $x=0.2$ (low noise) and $x=0.4$ (moderate noise). Figure~\ref{fig:fit_2q_rand} illustrates the results. The power-law fits align well with the data, as confirmed by the log-log residual diagnostics. For $x=0.2$, the extrapolated infinite-energy limit is $\Delta^\infty_E = -0.00017 \pm 0.00082$, approximately zero within uncertainty, which indicates near-ideal recovery of two-qubit coherences in the zero-noise limit. For $x=0.4$, the extrapolated value is $\Delta^\infty_E = 0.02888 \pm 0.00324$, revealing a small but non-zero residual error. This is consistent with the onset of threshold behavior observed in single-qubit analyses (Fig.~\ref{fig:extrapolation_over_x}), where moderate loss begins to overwhelm the code's corrective capacity, even at infinite energy.

These findings reinforce that ES-ZNE effectively mitigates finite-energy errors in multi-qubit settings, but they also highlight physical limitations: the non-zero extrapolation at higher loss suggests that the ideal GKP code cannot fully suppress decoherence beyond a certain noise threshold, motivating further investigation into optimized lattices or joint recovery maps.

\subsection{Quantifying the Computational Gain of ES-ZNE}

\begin{figure}[h!]
    \centering
    \includegraphics[width=\linewidth]{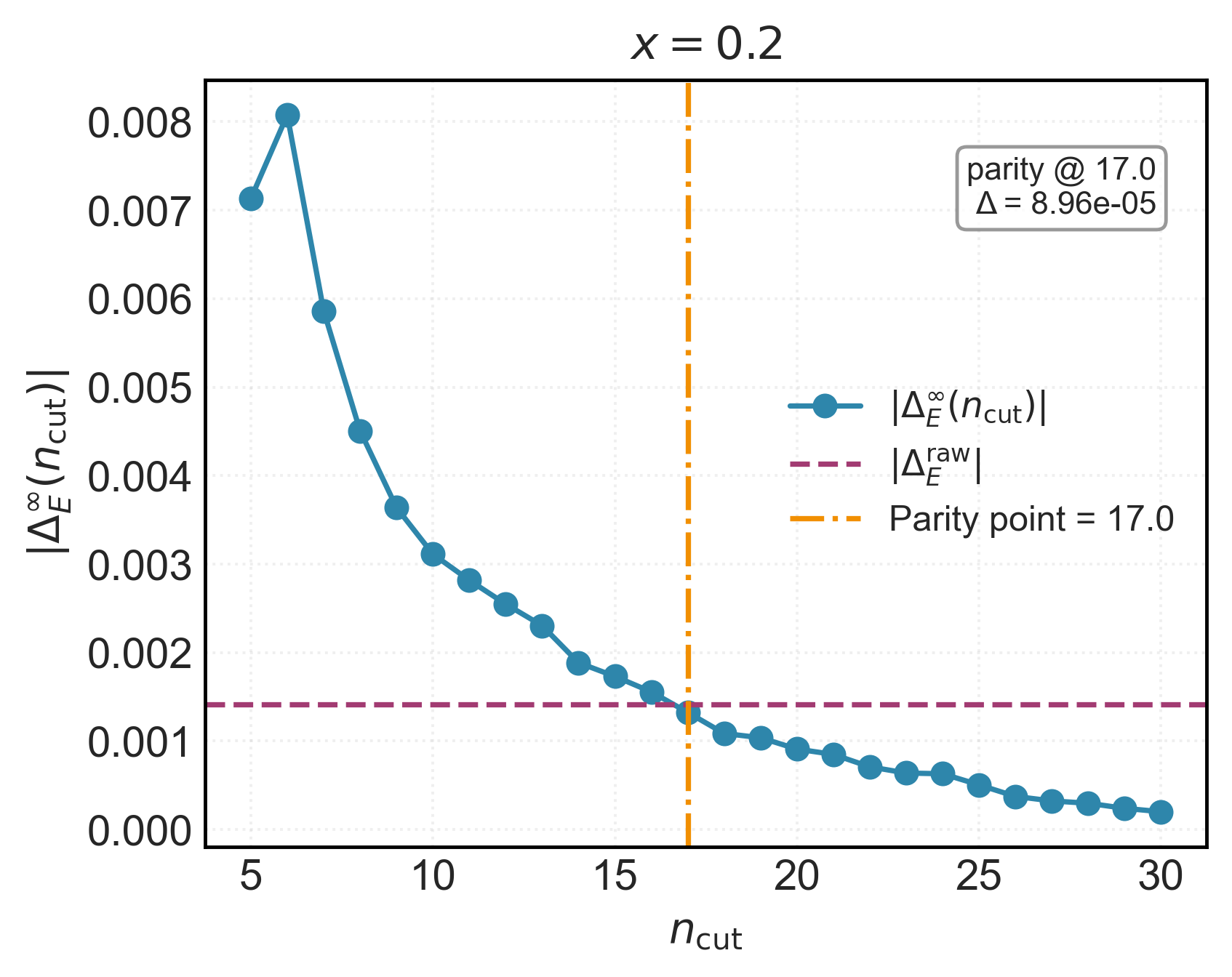} 
    \includegraphics[width=\linewidth]{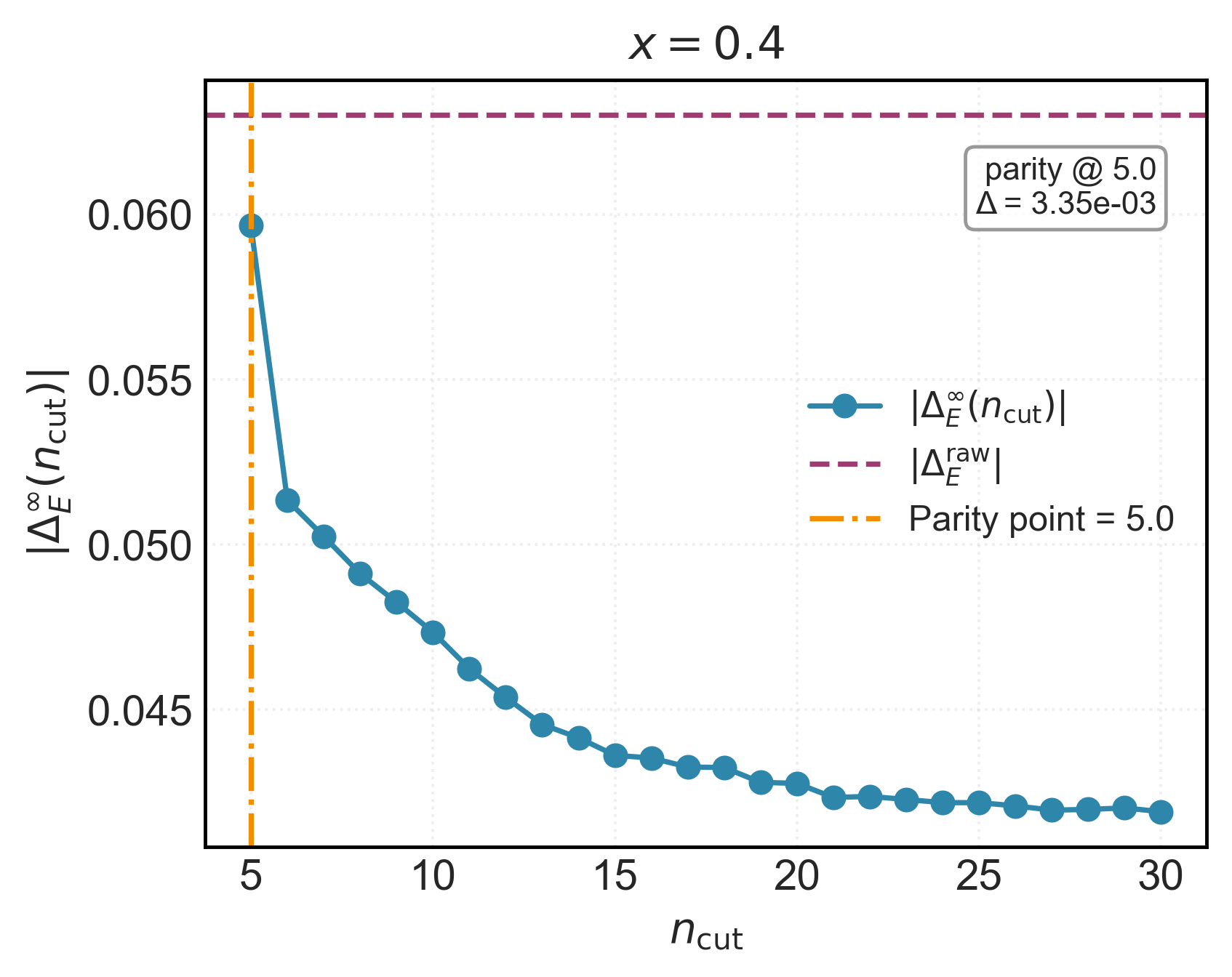} 
\caption{\textbf{Resource scaling and performance parity.} Absolute extrapolated error limit $|\Delta_E^{\infty}(n_{\text{cut}})|$ against the cutoff energy $n_{\text{cut}}$ used in the extrapolation. The dashed horizontal line indicates the raw benchmark error $|\Delta_E^{\text{raw}}|$ from the unmitigated simulation at $\bar{n}=30$. The vertical dash-dotted line marks the ``parity point'': the lowest $n_{\text{cut}}$ at which the extrapolated accuracy matches the raw high-energy benchmark. \textbf{(Top):} For $x=0.2$, parity is achieved with data up to $n_{\text{cut}} \approx 17$. \textbf{(Bottom):} For $x=0.4$, parity is achieved with data up to just $n_{\text{cut}} \approx 5$, indicating significant resource savings.}
    \label{fig:effective_nmax}
\end{figure}

Having demonstrated the effectiveness of ES-ZNE in mitigating finite-energy errors, we now quantify its practical utility by determining the computational resources saved. First, to validate the numerical stability of the protocol, we analyze the convergence of the extrapolated limit, $\Delta_E^{\infty}(n_{\text{cut}})$, as a function of the input data range. Instead of relying on a fixed dataset, we vary the maximum energy cutoff, $n_{\text{cut}}$, included in the fit (i.e., fitting only to data where $\bar{n} \le n_{\text{cut}}$). We observe that as $n_{\text{cut}}$ increases --- incorporating more data points closer to the asymptotic limit --- the extrapolated value converges to a stable solution.

With the stability of the method established, we quantify its computational advantage by determining the minimum experimental resources required to reproduce high-energy performance. We define a target benchmark, $\Delta_E^{\text{raw}}$, as the error metric measured directly at the highest available energy without extrapolation ($\bar{n}=30$). We then compare the extrapolated error limit $\Delta_E^{\infty}(n_{\text{cut}})$ derived from truncated datasets against this raw benchmark to identify the smallest $n_{\text{cut}}$ required to match the raw performance within numerical tolerance.

Figure~\ref{fig:effective_nmax} illustrates this analysis of performance parity. For the low-loss case ($x = 0.2$), we find that extrapolating from a dataset with a maximum energy of only $n_{\text{cut}} \approx 17$ is sufficient to recover the accuracy of the raw $\bar{n}=30$ simulation. The benefit is even more pronounced in the moderate-noise regime ($x = 0.4$), where extrapolating data up to just $n_{\text{cut}} \approx 5$ achieves the same result. These findings quantify the resource savings offered by ES-ZNE, demonstrating that a sequence of computationally inexpensive, low-energy simulations can effectively reproduce the precision of a much more demanding high-energy encoding.

\section{Conclusion}
\label{sec:conclusion}
In this work, we introduced Energy-Scaled Zero-Noise Extrapolation (ES-ZNE) as an instantiation of the zero-noise extrapolation method for Gottesman-Kitaev-Preskill (GKP) codes. Our approach is motivated by the substantial experimental challenges in directly improving the performance of bosonic codes. Achieving high intrinsic error resilience requires preparing high-energy GKP states, which in turn demands stringent hardware capabilities such as high squeezing levels, significant control overhead, and complex real-time feedback protocols \cite{Grimsmo2021, CampagneIbarcq2020}. Indeed, fault-tolerant architectures are predicted to require squeezing thresholds that are at the edge of current experimental capabilities \cite{NohChamberlandBrandao2022}. ES-ZNE provides an alternative software-based strategy. Instead of demanding better hardware to increase the code's energy, we leverage the principles of quantum error mitigation by trading temporal resources for improved accuracy. By systematically measuring observables at a series of lower, more accessible energy levels and extrapolating to the infinite-energy limit, our protocol aims to estimate the ideal performance without incurring the physical cost of preparing the ideal state, a foundational concept in error mitigation that exchanges sampling overhead for physical resources \cite{Temme2017, Cai2023, Takagi2023}.

By treating the mean photon number $\bar{n}$ as a tunable noise knob, ES-ZNE systematically mitigates finite-energy errors, whose monotonic scaling with energy is well-established both numerically \cite{Albert2018} and analytically \cite{Zheng2025}. We model this behavior with a power-law ansatz, $y(n) = L + c n^{-p}$, to extrapolate to the infinite-energy limit.   Our single-qubit simulations reveal a clear threshold behavior in the extrapolated performance: for low loss depths ($x \lesssim 0.3$), the code achieves near-perfect coherence, whereas a sharp "waterfall" degradation occurs around $x \approx 0.4$ (Fig.~\ref{fig:extrapolation_over_x}). This result was validated on two-qubit entangled states and through a coherence test with randomly-generated states, confirming the threshold as a fundamental code property.
This sharp performance degradation is a manifestation of a threshold effect, a general feature of quantum error correction. For GKP codes, this threshold is tied to a minimum required squeezing (energy) \cite{Fukui2018, Vuillot2019, Menicucci2014}, and similar transitions between protected and decoherence-dominated regimes are well-documented in both experimental and theoretical studies of bosonic codes under realistic constraints \cite{Ofek2016,Hu2019,Lescanne2020,CampagneIbarcq2020,Sivak2023,Royer2020,Noh2020,Hastrup2023,Shaw2024}.
The performance degradation above this threshold reflects the physical limitations of the code-recovery pair, where the complex, non-Pauli logical errors induced by photon loss can no longer be fully corrected \cite{Harris2025}.

Our results confirm that ES-ZNE is an effective error mitigation strategy for GKP codes under pure loss, but its utility is confined to the shallow-to-moderate noise regime associated with the protected side of the transition described above. In this regime, the protocol reliably recovers the ideal, error-free expectation values, whereas at larger loss depths its performance degrades and the extrapolation becomes increasingly uncertain. This behavior is a direct consequence of fundamental QEM sampling costs, which grow significantly as noise overwhelms the system and quantum states become indistinguishable \cite{Cai2023, Takagi2023}. By using a near-optimal Petz recovery, our extrapolation to the infinite-energy limit effectively isolates implementation-specific errors (finite energy) from the fundamental performance of the ideal code, whose limits align with analytic predictions \cite{ZhengPRL2024}. We therefore position ES-ZNE as a practical, software-based tool for enhancing the performance of near-term bosonic processors operating below their intrinsic error thresholds. Future work could extend this protocol to experimental data, explore alternative GKP lattices that may offer improved thresholds \cite{Hanggli2020}, or investigate its efficacy under correlated noise models. Ultimately, by leveraging computational time over physical hardware scale, ES-ZNE represents a valuable step toward practical quantum error correction in resource-constrained settings.

%%%%%%%%%%%%%%%%%%%%%%%%%
\section*{Acknowledgments}
This work was supported by
funding from the NSF QLCI for Hybrid Quantum Architectures and Networks (NSF award 2016136) and by the University of Wisconsin–-Madison L\&S Honors Program through The Trewartha Senior Honors Thesis Research Grant.

\bibliography{ref}

\end{document}